\documentclass[twocolumn,twoside]{IEEEtran}
\usepackage{mathpple}
\usepackage{times}
\usepackage{amsmath}  
\usepackage{amssymb}  
\usepackage{mathrsfs} 
\usepackage{theorem}  
\usepackage{cite}     
\usepackage{comment}  
\usepackage{upref}
\usepackage{amsfonts}
\usepackage{verbatim}
\usepackage[dvipsnames,usenames]{color}
\usepackage{enumerate}
\usepackage{cite}

\usepackage{caption}

\usepackage{algorithm}
\usepackage{times,color}
\usepackage{enumitem}
\usepackage{graphicx}
\usepackage{float}
\usepackage{color}
\usepackage{mathtools}
\usepackage{graphicx}
\usepackage{listings}
\usepackage{tikz}
\usepackage{multicol}
\usepackage{psfrag}
\usepackage{bbm}

\usepackage{pgfplots}
\usepackage{hhline}
\usepackage{xcolor,colortbl}
\usepackage{algorithm}
\usepackage[noend]{algpseudocode}

\algnewcommand{\Initialize}[1]{%
	\State \textbf{Initialize:}
	\Statex \hspace*{\algorithmicindent}\parbox[t]{.8\linewidth}{\raggedright #1}
}

\newcommand{\be}[1]{\begin{equation}\label{#1}}
\newcommand{\ee}{\end{equation}}

\newcommand{\bc}{\begin{center}}
\newcommand{\ec}{\end{center}}

\newcommand{\cC}{{\cal C}}

\newcommand{\cE}{{\cal E}}

\newcommand{\cG}{{\cal G}}
\newcommand{\cH}{{\cal H}}

\newcommand{\cS}{{\cal S}}

\newcommand{\bfc}{{\boldsymbol c}}

\newcommand{\bfv}{{\boldsymbol v}}
\newcommand{\bfw}{{\boldsymbol w}}
\newcommand{\bfx}{{\boldsymbol x}}
\newcommand{\bfy}{{\boldsymbol y}}
\newcommand{\bfz}{{\boldsymbol z}}
\newcommand{\bfA}{{\mathbf A}}

\newcommand{\bfI}{{\mathbf I}}

\renewcommand{\le}{\leqslant}
\renewcommand{\leq}{\leqslant}

\renewcommand{\geq}{\geqslant}

\newcommand{\Fq}{\smash{\mathbb{F}_{\!q}}}
\newcommand{\Ft}{\smash{\mathbb{F}_{\!2}}}
\newcommand{\Fqn}{\smash{\mathbb{F}_{\!q}^{\hspace{1pt}n}}}

\newcommand{\Ftn}{\smash{\mathbb{F}_{\!2}^{\hspace{1pt}n-1}}}

\newcommand{\Cref}[1]{Co\-rol\-la\-ry\,\ref{#1}}

\newcommand{\oin}{\smash{ \Sigma ^{n} }}

\newcommand{\sigqn}{\smash{\Sigma _q ^n}}
\newcommand{\sigqni}{\smash{\Sigma _q ^{n-1}}}

\theoremstyle{plain} \theorembodyfont{\normalfont\slshape}

\newtheorem{thm}{Theorem$\!$}
\newenvironment{theorem}{\begin{thm}\hspace*{-1ex}{\bf.}}{\end{thm}}

\newtheorem{prop}[thm]{Proposition$\!$}

\newtheorem{lem}[thm]{Lemma$\!$}
\newenvironment{lemma}{\begin{lem}\hspace*{-1ex}{\bf.}}{\end{lem}}

\newtheorem{cor}[thm]{Corollary$\!$}
\newenvironment{corollary}{\begin{cor}\hspace*{-1ex}{\bf.}}{\end{cor}}

\newtheorem{cons}[thm]{Construction$\!$}
\newenvironment{construction}{\begin{cons}\hspace*{-1ex}{\bf.}}{\end{cons}}

\newtheorem{defi}[thm]{Definition$\!$}

\newtheorem{cl}{Claim}
\newenvironment{claim}{\begin{cl}\hspace*{-1ex}{\bf .}}{\end{cl}}
\theorembodyfont{\normalfont}

\newtheorem{exam}{Example$\!$}

\newtheorem{remrk}{Remark$\!$}

\definecolor{Codecolor}{named}{White}  

\newcommand{\Copen}{\mbox{\{\kern-5.50pt\{}}
\newcommand{\Cclose}{\mbox{\}\kern-5.50pt\}}}
\newcommand{\Cslash}{\mbox{$\backslash\kern-6.02pt\backslash$}}

\newcommand{\wei}{\gamma}
\newcommand{\bwei}{\boldsymbol \gamma}

\newcommand{\etag}{\widehat{e}}

\newcommand{\shifted}{\cS}
\newcommand{\sho}{\cS_1}
\newcommand{\sht}{\cS_2}
\newcommand{\shr}{\cS_3}

\newcommand{\Sigq}{\Sigma_q}
\newcommand{\Sigt}{\Sigma_2}

\newcommand{\vto}{\ensuremath{\boldsymbol \alpha}}
\newcommand{\vtt}{\ensuremath{\boldsymbol \beta}}
\newcommand{\vtr}{\ensuremath{\boldsymbol \eta}}
\newcommand{\vtc}{\ensuremath{\mathbbm{1}}}

\newcommand{\cwx}{\ensuremath{\bfx}}
\newcommand{\cwy}{\ensuremath{\bfy}}
\newcommand{\cww}{\ensuremath{\bfz}}

\newcommand{\sot}{\sum_{j=1}}

\begin{document}
\title{\textbf{Single-Deletion Single-Substitution \\Correcting Codes}}

\author{%
	\IEEEauthorblockN{\textbf{Ilia Smagloy$^1$}\IEEEauthorrefmark{1},
                \textbf{Lorenz Welter$^1$}\IEEEauthorrefmark{2},
                \textbf{Antonia Wachter-Zeh}\IEEEauthorrefmark{2},
		and \textbf{Eitan Yaakobi}\IEEEauthorrefmark{1},} \\
        \IEEEauthorblockA{\IEEEauthorrefmark{1}%
		Technion --- Israel Institute of Technology, Israel,
		\{ilia.smagloy, yaakobi\}@cs.technion.ac.il} \\
        \IEEEauthorblockA{\IEEEauthorrefmark{2}%
		Technical University of Munich, Germany,
		\{lorenz.welter, antonia.wachter-zeh\}@tum.de} 
	
	\thanks{$^1$The first two authors contributed equally to this work. The work of L. Welter and A. Wachter-Zeh has received funding from the European Research Council (ERC) under the European Union’s Horizon 2020 research and innovation programme (grant agreement No. 801434). The work of I. Smagloy and E. Yaakobi was partially supported by the United States-Israel BSF grant 2018048.}\vspace{-4ex}
}

\maketitle

\begin{abstract}
Correcting insertions/deletions as well as substitution errors simultaneously plays an important role in DNA-based storage systems as well as in classical communications.
This paper deals with the fundamental task of constructing codes that can correct a single insertion or deletion along with a single substitution.
A non-asymptotic upper bound on the size of single-deletion single-substitution correcting codes is derived, showing that the redundancy of such a code of length $n$ has to be at least $2 \log n$. The bound is presented both for binary and non-binary codes while an extension to single deletion and multiple substitutions is presented for binary codes. An explicit construction of single-deletion single-substitution correcting codes with at most $6 \log n + 8$ redundancy bits is derived. Note that the best known construction for this problem has to use 3-deletion correcting codes whose best known redundancy is roughly $24 \log n$.
\end{abstract}

\section{Introduction} \label{sec:intro}
Codes correcting insertions/deletions recently attract a lot of attention due to their relevance in DNA-based data storage systems, cf. \cite{Heckel_A-Characterization-of-the-DNA-Storage-Channel_19}. In classical communications, insertions/deletions happen during the synchronization of files and symbols of data streams \cite{SalaSchoeny-Sync_TransComm} or due to over-sampling and under-sampling at the receiver side~\cite{DolecekAnan_Sync_2007}. 
The algebraic concepts correcting insertions and deletions date back to the 1960s when Varshamov and Tenengolts designed a class of binary codes, nowadays called \emph{VT codes}. These codes were originally designed to correct asymmetric errors in the $Z$-channel~\cite{tenengolts1984nonbinary,VarshTene-SingleDeletion1965} and later proven to be able to correct a single insertion or a single deletion~\cite{Levenshtein-binarycodesCorrectingDeletions}. 
VT codes are asymptotically optimal length-$n$ single-insertion/deletion correcting codes of redundancy $\log(n+1)$. 
As a generalization of VT codes, Tenengolts presented $q$-ary single-insertion/deletion correcting codes in~\cite{tenengolts1984nonbinary}. Levenshtein has also proven that for correcting $t$ insertions/deletions, the redundancy is asymptotically at least $t \log n$.
In~\cite{BrakensiekGuruswamiZbarksy-DeletionCorrection}, Brakensiek \emph{et al.} presented binary multiple-insertion/deletion correcting codes with small asymptotic redundancy.
For an explicit small number of deletions, their construction however needs redundancy $c \log n$ where $c$ is a large constant.
The recent parallel works by Gabrys \emph{et al.} \cite{Gabrys-TwoDeletions_2018} and Sima \emph{et al.} \cite{Sima-TwoDeletions_2019} have presented constructions to correct two deletions with redundancy $8 \log n + O (
\log \log n)$ \cite{Gabrys-TwoDeletions_2018} and $7\log n + o(\log n)$ \cite{Sima-TwoDeletions_2019}, respectively. Sima and Bruck \cite{SimaBruck-kDeletions_2019} generalized their construction to correct any $t$ insertions/deletions with redundancy $8t\log n + o(\log n)$.

However, in DNA data storage as well as in file/symbol synchronization, not only insertions/deletions occur, but also classical substitution errors.
Clearly, a substitution error can be seen as a deletion followed by an insertion. Therefore, in order to correct for example a single deletion and a single substitution, the best known construction uses codes correcting \emph{three} deletions. The leading construction for three-deletion correcting codes is the one by Sima and Bruck \cite{SimaBruck-kDeletions_2019} which has redundancy roughly $24 \log n$.

In this paper, we initiate the study of codes correcting substitutions and insertions/deletions. One of our main results is a construction of a single-deletion single-substitution correcting code 
with redundancy at most $6 \log n + 8$ which significantly improves upon using 3-deletion correcting codes. We also derive a non-asymptotic upper bound on the cardinality of $q$-ary single-deletion single-substitution codes which shows that at least redundancy $2 \log n$ is necessary (in contrast to a 3-deletion correcting code that requires redundancy at least $3 \log n$). In the binary case this bound is also generalized to single-deletion multiple-substitution correcting codes.


\section{Definitions and Preliminaries}\label{sec:def}
This section formally defines the codes and notations that will be used throughout the paper. For two integers $ i, j \in \mathbb{N} $ such that $ i \le j $ the set $ \{i, i+1, \ldots, j\} $ is denoted by $ [i,j] $ and in short $[j]$ if $i=0$. The alphabet of size $q$ is denoted by $\Sigma_q=\{0,1,\ldots,q-1\}$. A $t$-$indel$ is any combination of $t_D$ deletions and $t_I$ insertions such that $t_D + t_I = t$.  Moreover, for two positive integers, $t \leq n$ and $s \leq n-t$, $B^{DS}_{t,s}(\bfx)$ is the set of all words received from $\bfx \in \sigqn$ after $t$ deletions and at most $s$ substitutions. Note that the order in which the errors occur does not matter and thus we will mostly assume that first the deletions occurred. Finally, $r(\bfx)$ denotes the number of runs in $\bfx \in \sigqn $. 

A code $\cC\subseteq \sigqn$ is called a \emph{$t$-deletion $s$-substitution correcting code} if it can correct any combination of at most $t$ deletions and $s$ substitutions. That is, for all $\bfc_1,\bfc_2\in\cC$ it holds that $B^{DS}_{t,s}(\bfc_1) \cap B^{DS}_{t,s}(\bfc_2) =\emptyset$. We define similarly \emph{$t$-indel $s$-substitution correcting code} to be a code that corrects any combination of at most $t$ indels and $s$ substitutions.

The goal of this paper is to study codes correcting indels and substitutions. Similarly to the equivalence between insertion and deletion correcting codes, the following lemma holds.

\begin{lemma}
	A code $\cC$ is a $t$-indel $s$-substitution correcting code if and only if it is a $t$-deletion $s$-substitution-correcting code. 
\end{lemma}

Therefore, the main focus of the paper is on $t$-deletion $s$-substitution correcting codes and specifically for $t=1$. The size of the largest $q$-ary length-$n$ single-deletion $s$-substitution correcting code is denoted by $DS_{s,q}(n)$.


\section{Bounds}\label{sec:bounds}
The method used to compute a non-asymptotic upper bound for the cardinality of any single-deletion $s$-substitution code is described in \cite{Kulkarni_Nonasymptotic-Upper-Bounds-for-Deletion-Correcting-Codes_13} and \cite{Fazeli_Generlized-Sphere-Packing-Bound_15}. For clarity of the results, the principal concepts of this method are briefly reviewed. The main idea is to construct a hypergraph $\cH _s(X,\cE_s)$ out of the channel graph with vertices $X = \sigqni = \{\bfx_1,\dots,\bfx _m \}$ and hyperedges $\cE_s = \{E_1, \dots, E_\ell \} = \{B^{DS}_{1,s}(\bfx) : \bfx \in \sigqn \}$. The objective is to find the smallest size of a transversal $T \subseteq X$ in $\cH_s$, i.e. $T$ intersects all hyperedges in $\cH_s$. Let $\bfI$ be the $m \times \ell$ incidence matrix of $\cH$ where $\bfI (i,j) = 1$ if $\bfx_i \in E_j$. A transversal $\bfw\in\Sigma_2^m$ satisfies that $\bfI ^T \cdot \bfw \geq 1$. If $\bfw \in (\mathbb{R}^+)^m$, then it is called a fractional transversal. Thus, the objective is to find some $w_{\bfy} \geq 0$, which needs to fulfill the condition $\sum _{\bfy \in B^{DS}_{1,s}(\bfx) } w_{\bfy} \geq  1$ for all $\bfx \in \sigqn$. Consequently, the following expression is an upper bound of the cardinality of a code,
\begin{equation*}
\vert \cC \vert \leq \sum \nolimits _{\bfy \in \sigqni} w_{\bfy}. \label{eq:bound-def}
\end{equation*}

\vspace{-2ex}
\subsection{Upper Bound on Single-Deletion Single-Substitution Codes}
Before determining valid fractional transversals, an important property for any $\bfy \in B^{DS}_{1,s}(\bfx)$ is studied in the following.
\vspace{-3.5ex}
\begin{claim} \label{claim:runchange}
	For all  $\bfx \in \sigqn$ and $\bfy \in B^{DS}_{1,s}(\bfx)$, it holds that 
	\begin{equation*}
	r(\bfx) - (2+2s) \leq r(\bfy) \leq r(\bfx) + 2s. \label{eq:run-x-to-y}
	\end{equation*}
\end{claim}
Thus, the monotonicity argument $\vert  B^{DS}_{1,s}(\bfy) \vert \leq \vert B^{DS}_{1,s}(\bfx) \vert$ as in the single deletion case of \cite{Kulkarni_Nonasymptotic-Upper-Bounds-for-Deletion-Correcting-Codes_13} does not necessarily hold and choosing $w_{\bfy} = \frac{1}{\vert B^{DS}_{1,s}(\bfy)\vert}$ does not suffice as  a feasible solution.  

The following lemma extends the result from~\cite{AbuSini_Reconstruction-of-Sequences-in-DNA-Storage_19} and provides the size of $B^{DS}_{1,1}(\bfx)$ for the $q$-ary alphabet.
\begin{lemma} \label{lem:ballsize-q}
For any word $\bfx \in \sigqn$, 
\begin{flalign*}
\vert B^{DS}_{1,1}(\bfx) \vert = 
\begin{cases}
(n-1)(q-1) + 1 \hspace{12ex}   r(\bfx) = 1, \\
r(\bfx) \left[ (n-3) (q-1) + (q-2) \right] + (q+2) \\ \hspace{30ex} r(\bfx) \geq 2.
\end{cases} \label{eq:ball-size}
\end{flalign*}
\end{lemma}

Now, from the results of Lemma \ref{lem:ballsize-q} and Claim \ref{claim:runchange}, a valid expression of a fractional transversal $w_{\bfy}$ can be derived.

\begin{lemma} \label{lemma:qeq2}
	The following choice of $w_{\bfy}$ for $\bfy \in \sigqni, n \geq 3$, is a fractional transversal for $\cH_1$
	\begin{flalign*}
	w_{\bfy}^1 = 
	\begin{cases}
	\frac{1}{(n-1)(q-1)+1} &\quad r(\bfy) \leq 3 \\
	\frac{1}{(r(\bfy)-2) \left[ (n-3) (q-1) + (q-2) \right] + (q+2)} &\quad r(\bfy) > 3 .
	\end{cases} 
	\end{flalign*}
\end{lemma}

The following claim will be used in computing the upper bound of the cardinality of the code.
\vspace{-1ex}
\begin{claim} \label{claim:binom-r-upper-bound1}
	For integers $q \geq 2$, $n \geq 5$, and $n \geq q$ it holds	that 
	\begin{equation*}
	\sum _{k=1} ^n \binom{n}{k} (q-1)^k \frac{1}{k} \leq \frac{q^{n+1}}{(q-1)(n-2)}.
	\end{equation*}
\end{claim}

Note that the claim is combined from similar statements in \cite[Lemma 13, 14]{Gabrys_Correcting-Grain_Errors-in-Magnetic-Media_15}. Putting everything together the following upper bound on $DS_{1,q}(n)$ can be presented. 

\begin{theorem} \label{theorem:nonbinarybound}
For $q\leq n, n \geq 6$ the following is an upper bound on $DS_{1,q}(n)$
\vspace{-1ex}
	\begin{flalign*}
	DS_{1,q}(n) &\leq \frac{3 \cdot q^{n-1}}{(n-5)(n-3)(q-1)} + 5q
	\end{flalign*}	
\end{theorem}
\begin{IEEEproof}
Note that the number of words in $\Sigma_q^{n-1}$ with $r$ runs is given by $q(q-1)^{r-1}\binom{n-2}{r-1}$~\cite{Kulkarni_Nonasymptotic-Upper-Bounds-for-Deletion-Correcting-Codes_13}. The sum over all words in $\sigqni$ using the indicated fractional transversals $w_{\bfy}^1$ has to be computed. For $r = 1, 2, 3$ define the function $g(q,n) = \sum _{r=1} ^3 q(q-1)^{r-1} \binom{n-2}{r-1} w_{\bfy}^1$.  The rest is given by
{\small
\begin{flalign*}
&\sum _{r=4} ^{n-1} q (q-1)^{r-1} \tbinom{n-2}{r-1} \frac{1 }{(r-2) \left[ (n-3) (q-1) + (q-2) \right] + (q+2)} \\
&\leq \frac{q}{(n-3)(q-1)} \sum _{r=4} ^{n-1} (q-1)^{r-1} \binom{n-2}{r-1} \frac{1}{r-2}. 
\end{flalign*}
}
For simplicity, first the following analysis is performed.
{\small
\begin{flalign*}
&f(q,n) \coloneqq \sum _{r=4} ^{n-1} (q-1)^{r-1} \binom{n-2}{r-1} \frac{1}{r-2} \\
&= \sum _{r=2} ^{n-3} (q-1)^{r+1} \frac{(n-2)!}{(n-r-3)!\,r!} \left( \frac{1}{r} - \frac{1}{r+1} \right) \\
&=(n-2) (q-1) \sum _{r=2} ^{n-3} (q-1)^{r} \binom{n-3}{r} \frac{1}{r}  - \sum _{r=2} ^{n-3} (q-1)^{r+1} \binom{n-2}{r+1}.
\end{flalign*}
}
In the last expression the right part of the difference can be calculated as follows
{\small
\begin{flalign*}
&\sum _{r=2} ^{n-3} (q-1)^{r+1} \binom{n-2}{r+1} = \sum _{r=3} ^{n-2} (q-1)^{r} \binom{n-2}{r}\\ 
&=  q ^{n-2} - \frac{(q-1)^2(n-2)(n-3)}{2} - (q-1)(n-2) - 1 .
\end{flalign*}
}
For the left part, Claim \ref{claim:binom-r-upper-bound1} can be used to derive the following inequality
{\small
\begin{flalign*}
(n-2) & (q-1)\sum _{r=2} ^{n-3} (q-1)^{r} \binom{n-3}{r} \frac{1}{r} \\ 
&\leq (n-2)(q-1) \left[ \frac{q^{n-2}}{(q-1)(n-5)} - (n-3)(q-1) \right] .
\end{flalign*}
}
Thus, an upper bound for $f(q,n)$ can be derived by
{\small
\begin{flalign*}
f(q,n) &\leq \left[ \frac{(n-2)}{(n-5)}-1 \right] q^{n-2} \\
& \quad - \frac{1}{2} (n-2)(n-3)(q-1)^2 + (q-1) (n-2) + 1.
\end{flalign*}
}
Next, the computed $f(q,n)$ and $g(q,n)$ are combined in the following manner  
\begin{flalign*}
DS_{1,q}(n) &\leq \frac{q \cdot f(q,n)}{(n-3)(q-1)} + g(q,n).
\end{flalign*}
Finally, the bound in the theorem results after some basic algebraic steps and the fact that $q \leq n$.
\end{IEEEproof}
The last theorem provides the following corollary. 
\begin{corollary} \label{cor:bound-q-1}
It holds that $DS_{1,q}(n)\lesssim\frac{3\cdot q^{n-1}}{n^2(q-1)}.$
\end{corollary}

\subsection{Upper Bound on Single-Deletion s-Substitution Codes}
To state a legitimate fractional transversal for the case of $s$ substitutions, first a lower bound on the cardinality of the ball size $\lvert B ^{DS} _{1,s}(\bfx) \rvert$ has to be derived. In the remaining part of the section only $\Sigt$ will be considered.
\vspace{-1ex}
\begin{claim} \label{claim:lower-bound-s-subs-ball}
For all $\bfx \in \Sigma_2^n$, it holds that $ \lvert B ^{DS}_{1,s}(\bfx) \rvert \geq  r(\bfx) \binom{n-1-s}{s}$.
\end{claim} \vspace{-1ex}

Note that this lower bound is derived based upon an explicit expression of $\lvert B ^{DS} _{1,s}(\bfx) \rvert$ from \cite{AbuSini_Reconstruction-of-Sequences-in-DNA-Storage_20}.
Using this result, a fractional transversal for the single-deletion $s$-substitution case can be formulated. 
\begin{lemma} \label{lem:frac-s-sub}
The following choice of $w_{\bfy}$ with $\bfy \in  \Sigma_2^{n-1}$ and $n \geq 2s + 1 \geq 3$ is a fractional transversal for $\cH_s$
\begin{flalign*}
w_{\bfy} ^s= 
\begin{cases}
\frac{1}{\binom{n-s-1}{s}} &\quad r(\bfy) \leq 2s+1, \\
\frac{1}{(r(\bfy)-2s) \binom{n-s-1}{s}} &\quad r(\bfy) > 2s+1 .
\end{cases} \label{eq:transversal-1-s-q2}
\end{flalign*}
\end{lemma}
As a result of Lemma \ref{lem:frac-s-sub} an upper bound for the cardinality of a single-deletion $s$-substitution correcting code can be stated. 
\begin{theorem} \label{theorem:ssubsbinary}
For $n \geq 3$ the following is an upper bound on $DS_{s,2}(n)$: \vspace{-2ex}
{\small
\begin{flalign*}
DS_{s,2}(n) &\leq \frac{s!(2s+1)}{(n-2s)^{s}(n-1)} \left[ 2^n + \frac{2 (n-1)^{2s+1}}{2s+1} \right].
\end{flalign*}
}
\end{theorem}
\begin{IEEEproof}
First, only the words in $\bfy \in \Sigma_2^{n-1}$ with $r(\bfy) \leq 2s+1$ are considered. Using the inequalities are $ \sum _{i=0}^{k}{n \choose i}\leq \sum _{i=0}^{k}n^{i}\cdot 1^{k-i}\leq (1+n)^{k}$ and $\binom{n}{k} \geq \frac{(n-k+1)^k}{k!}$, the sum can be calculated as follows
{\small
\begin{flalign*} 
\sum ^{2s+1} _{r=1} 2 \binom{n-2}{r-1} \frac{1}{\binom{n-s-1}{s}} = \frac{2}{ \binom{n-s-1}{s}} \sum ^{2s} _{r=0} \binom{n-2}{r} \leq \frac{2 s! (n-1) ^ {2s} }{(n-2s)^s}. 
\end{flalign*}
}
In a next step, by additionally applying the inequality $\frac{1}{r-2s} \leq \frac{2s+1}{r}$ the sum for all words with $2s+2 \leq r(\bfy) \leq n-1$ can be computed as
{\small
\begin{flalign*} 
&\sum ^{n-1} _{r=2s+2} 2 \binom{n-2}{r-1} \frac{1}{ \binom{n-s-1}{s}} \frac{1}{r-2s} \leq \frac{2}{ \binom{n-s-1}{s}} \sum ^{n-1} _{r=2s+2} \binom{n-2}{r-1} \frac{2s+1}{r} \nonumber \\
&= \frac {2s+1}{n-1} \frac{2}{ \binom{n-s-1}{s}} \sum ^{n-1} _{r=2s+2} \binom{n-1}{r} \leq \frac {2s+1}{n-1} \frac{2 s!}{(n-2s)^s}\cdot 2^{n-1}. \nonumber 
\end{flalign*}
}
Subsequently, the sum of all words with $r \leq 2s+1$ is added to the equation again which results to the expression in the theorem.
\end{IEEEproof}
The corollary below concludes the previous result. 
\begin{corollary} \label{cor:ssubs}
It holds that $DS_{s,2}(n) \lesssim \frac{s! (2s+1) \cdot 2^{n}}{n^{s+1}}$. 
\end{corollary}

Note that unlike the proof of Theorem \ref{theorem:nonbinarybound}, in the proof of Theorem \ref{theorem:ssubsbinary} Claim \ref{claim:binom-r-upper-bound1} is not applied. Instead, Claim \ref{claim:lower-bound-s-subs-ball} and a different upper bound is used. For this reason, in Corollary \ref{cor:ssubs} the bound for the value of $s=1$ is not the same as the bound stated in Corollary \ref{cor:bound-q-1} with $q=2$.

\section{Properties of Codes}\label{sec:prop}
In this section, several properties of the families of codes studied in the paper are presented. Consider a family of codes which are defined in the following way. A binary  code $\cC\subseteq \Sigt^n$ is called a \emph{$(\bwei = (\wei_1,\ldots,\wei_n);a,N)$-congruent code} if it is defined in the following way
$$\cC(\bwei;a,N) = \left\{ \bfc\in \Sigt^n \ | \ \sum_{i=1}^n \wei_i\cdot c_i \equiv a\, \ (\bmod N) \right\}.$$
The lemmas in this section provide several basic properties in case the intersection of the single-deletion single-substitution balls of two codewords in $\cC$ is not trivial. For the rest of this section it is assumed that $\cwx,\cwy\in \cC(\bwei;a,N)$, where $B^{DS}_{1,1}(\cwx) \cap B^{DS}_{1,1}(\cwy) \neq \emptyset$ so that there exists $\cww \in B^{DS}_{1,1}(\cwx) \cap B^{DS}_{1,1}(\cwy)$. 
For simplicity of notation, the expression $\cwx(d,e)$ is defined to be the error-word achieved from $\cwx$ by deleting the bit in the index $d$, and substituting the bit in the index $e$. 
The variables $d_x,d_y,e_x,e_y\in [n]$ are indices such that $\cww=\cwx(d_x,e_x)=\cwy(d_y,e_y)$.
It can be assumed w.l.o.g. that $d_x < d_y$. In order to shift the values of the substituted bits from binary to $\pm 1$, the following notation is used $\delta_i:=2\cdot x_i-1$. 
\begin{lemma} \label{lemma:exouteyout}
For $e_x,e_y\notin [d_x,d_y]$ the following statements hold.
\begin{enumerate}
\item For $i\in [d_x+1,d_y]$, $x_i = y_{i-1}$.
\item For $i \in \{e_x, e_y\}$,  $x_i = \overline{y_i}=1-y_i$.
\item For $i\in [n]\setminus([d_x,d_y]\cup \{e_x, e_y\})$, $x_i = y_i$.
\item $\sum_{i=d_x}^{d_y}\wei_i\cdot(x_i-y_i) =$ \\$x_{d_x}\cdot\wei_{d_x} + \sum_{i=d_x+1}^{d_y} (\wei_i - \wei_{i-1})\cdot x_i - y_{d_y}\cdot\wei_{d_y}$
\item 
$	\delta_{e_x} \cdot \wei_{e_x} + \delta_{e_y}\cdot \wei_{e_y} + x_{d_x}\cdot \wei_{d_x} - y_{d_y}\cdot \wei_{d_y} \\
	+\sum_{i=d_x+1}^{d_y} x_i\cdot(\wei_i-\wei_{i-1}) \equiv 0 \ (\bmod N)
$
\end{enumerate}
\end{lemma}
\begin{IEEEproof}
For any $i\in [d_x+1,d_y]$ the definition of $\cww$ leads to the fact that $z_i=y_i$ and also $z_i=x_{i+1}$. This proves statement $1$, and a similar proof can be shown for statement $3$.

For a substituted bit $i=e_x$ either $i<d_x$ in which case, $x_i=\overline{z_i}, y_i=z_i$. 
The cases of $i= e_y$ and $i>d_y$ can be proved in a similar way. This concludes the proof of statement $2$. 

In order to prove statement 4, the  sum is separated in the following manner\vspace{-1ex}
{\small
	$$\sum_{i=d_x}^{d_y}\wei_i\cdot(x_i-y_i) =x_{d_x}\cdot\wei_{d_x}- y_{d_y}\cdot\wei_{d_y}  + \sum_{i=d_x+1}^{d_y}\wei_i\cdot x_i - \sum_{i=d_x}^{d_y-1}\wei_i\cdot y_i.$$
}
Using statement $1$, the last element is simplified as follows\vspace{-1ex}
$$\sum_{i=d_x}^{d_y-1}\wei_{i}\cdot y_i = \sum_{i=d_x+1}^{d_y}\wei_{i-1}\cdot y_{i-1} = \sum_{i=d_x+1}^{d_y}\wei_{i-1}\cdot x_{i}.$$ 
Thus, the equality can be rewritten as \vspace{-3ex}

{\small
\begin{align*}
&x_{d_x}\cdot \wei_{d_x} - y_{d_y}\cdot \wei_{d_y} + \sum_{i=d_x+1}^{d_y}\wei_i\cdot x_i - \sum_{i=d_x}^{d_y-1}\wei_i\cdot y_i & \\
&= x_{d_x}\cdot \wei_{d_x} - y_{d_y}\cdot \wei_{d_y} +\sum_{i=d_x+1}^{d_y}(\wei_i-\wei_{i-1})\cdot x_{i}.& 
\end{align*}
}
This concludes the proof of statement $4$.\par

The fact $\cwx,\cwy\in \cC(\bwei;a,N)$ implies that $\sum_{i=1}^n \wei_ix_i \equiv \sum_{i=1}^n \wei_iy_i \equiv a \ (\bmod N)$. From this follows \vspace{-1ex}
$$ \sum_{i=1}^n \wei_ix_i - \sum_{i=1}^n \wei_iy_i\equiv 0 \ (\bmod N). $$
Furthermore, Statement $3$ implicates that 
{\small$\sum_{i=1}^{d_x-1} \wei_i\cdot(x_i-y_i) +  $\newline$
\sum_{i=d_y+1}^{n} \wei_i\cdot(x_i-y_i) = \wei_{e_x}\cdot(x_{e_x}-y_{e_x}) + \wei_{e_y}\cdot(x_{e_y}-y_{e_y})$}. On the other hand statement $2$ leads to the fact that for $e\in \{e_x,e_y\}: x_e-y_e = x_e-\overline{x_e} = 
2\cdot x_e-1=\delta_{x_e}$. Combined together with statement $4$ the following equivalence is achieved
\vspace{-1ex}
{\small
\begin{flalign*}
\delta_{e_x} \cdot \wei_{e_x} + \delta_{e_y}\cdot \wei_{e_y} &+ x_{d_x}\cdot \wei_{d_x} +\sum_{i=d_x+1}^{d_y} x_i\cdot(\wei_i-\wei_{i-1})- y_{d_y}\cdot \wei_{d_y} \\
&= \sum_{i=1}^n \wei_ix_i - \sum_{i=1}^n \wei_iy_i \equiv 0 \ (\bmod N).
\end{flalign*}
}
This proves statement $5$.
\end{IEEEproof}

In a similar manner, the following lemma can be proved.\vspace{-2ex}

\begin{lemma} \label{lemma:exeycases}
The following conditions hold:
\begin{enumerate}
\item For $e_x\in [d_x,d_y]$ and $e_y\notin [d_x,d_y]$
\begin{center}
$ \delta_{e_x} \cdot \wei_{e_x-1} + \delta_{e_y}\cdot \wei_{e_y} + x_{d_x}\cdot \wei_{d_x}- y_{d_y}\cdot \wei_{d_y}+$
$+\sum_{i=d_x+1}^{d_y} x_i\cdot(\wei_i-\wei_{i-1})
 \equiv 0 \ (\bmod N)$
\end{center}
\item For $e_x,e_y\in [d_x,d_y]$
\begin{center}
$ \delta_{e_x} \cdot \wei_{e_x-1} + \delta_{e_y+1}\cdot \wei_{e_y} + x_{d_x}\cdot \wei_{d_x}- y_{d_y}\cdot \wei_{d_y}+$
$+\sum_{i=d_x+1}^{d_y} x_i\cdot(\wei_i-\wei_{i-1})
 \equiv 0 \ (\bmod N)$

\end{center}
\end{enumerate}
\end{lemma}

Define the variable $\epsilon_x$ as 1 if $e_x\in [d_x,d_y]$ and 0 otherwise. 
The variable $\epsilon_y$ is defined in a similar manner.
An important claim about the property of the code is as follows. \vspace{-2ex}

\begin{claim} \label{claim:complprop}
For any $\cwx,\cwy\in \Sigt^n$
\begin{enumerate}
\item {\small$\hspace{0.03pt}\cwx(d_x,e_x)\hspace{0.03pt}=\hspace{0.03pt}\cwy(d_y,e_y)$ iff $\hspace{0.03pt}\cwx(d_x,e_y+\epsilon_y)\hspace{0.03pt}=\hspace{0.03pt}\cwy(d_y,e_x-\epsilon_x)$.}
\item $B_{1,1}(\cwx)\cap B_{1,1}(\cwy)\neq \emptyset$ iff $B_{1,1}(\overline{\cwx})\cap B_{1,1}(\overline{\cwy})\neq \emptyset$.
\item For some weight vector $\bwei\in\mathbb{Z}^n,\ a, N$, and for any $\cwx,\cwy\in \cC(\bwei;a,N)$ there exists such $b$ so that $\overline{\cwx}, \overline{\cwy} \in \cC(\bwei;b,N)$.
\end{enumerate}
\end{claim}

\section{Construction}\label{sec:cons}
In this section, the main result of the paper is shown. An explicit construction for a single-deletion single-substitution correcting code and its correctness are presented. This construction requires redundancy of at most $6\cdot \log (n) +8$ bits.

\begin{construction}
Define four weight vectors $\vto,\vtt, \vtr, \vtc \in\mathbb{Z}^n$ in the following way\vspace{-1ex}
$$\vto = (1,2,3,\ldots,n), \vtt=(\sum_{i=1}^1i,\sum_{i=1}^2i,\sum_{i=1}^3i,\ldots,\sum_{i=1}^ni),$$ $$ \vtr =(\sum_{i=1}^1i^2,\sum_{i=1}^2i^2,\sum_{i=1}^3i^2,\ldots,\sum_{i=1}^ni^2),  \vtc=(1,\ldots ,1).$$ For fixed integers $a\in [3n],b\in [3\cdot n^2],c\in [3\cdot n^3],d\in [4]$, the code is defined as
\begin{flalign*}
\cC_{a,b,c,d} &= \cC(\vto;a,3n+1) \cap \cC(\vtt;b,3n^2+1) \\
& \cap \, \cC(\vtr;c,3n^3+1) \cap \cC(\vtc;d,5)
\end{flalign*}
\end{construction}

Remember that $d_x,d_y$ are denoted as the indices of the deleted bits from $\cwx,\cwy$ respectively. The definition of $e_x,e_y$ are the indices of the substituted bits from $\cwx,\cwy$.
The definition of $\epsilon_x$ is 1 if $e_x\in[d_x,d_y]$ and 0 otherwise. The definition of $\epsilon_y$ is similar.\vspace{-2ex}

\begin{theorem}  \label{theorem:onedelondesub}
For any indices $e_x,e_y,d_x,d_y\in [n]$, any two codewords $\cwx,\cwy\in \cC_{a,b,c,d}$ fulfill $$\cwx(d_x,e_x)\neq \cwy(d_y,e_y).$$ 
\end{theorem}

\begin{IEEEproof} For simplicity, denote $\etag_x = e_x-\epsilon_x$ and $\etag_y=e_y+\epsilon_y$.
Assume by contradiction that $\cwx(d_x,e_x)=\cwy(d_y,e_y)$. The following equivalence can be concluded from Lemma \ref{lemma:exouteyout} (statement 5), Lemma \ref{lemma:exeycases}, and the fact that $\cC_{a,b,c,d}\subseteq \cC(\vtc;d,5)$.
$$
\delta_{e_x} + \delta_{\etag_y} + x_{d_x} - y_{d_y} + 0 \equiv 0 \bmod 5
$$
This is equivalent to the following system of equations
$$
x_{d_x}=y_{d_y},\ \delta_{e_x}=-\delta_{\etag_y}.
$$

Define the set $\shifted$ as follows. $$\shifted \coloneqq \{d_x+1\leq i\leq d_y| x_i=1\}\subseteq [d_x,d_y].$$ 
According to Claim  \ref{claim:complprop} statement 1, it is possible to assume w.l.o.g. that $e_x<e_y$.
According to Claim  \ref{claim:complprop} statements 2-3, it is also possible to assume w.l.o.g. that $x_{d_x}=y_{d_y}=0$. 
By substituting these values into Lemma \ref{lemma:exouteyout} statement 5 and Lemma \ref{lemma:exeycases} the following equivalence is achieved.

For any $k \in \{0,1,2\}$
$$\delta_{e_x}\cdot \sot^{\etag_x}j^k + \delta_{\etag_y}\cdot \sot^{e_y}j^k+ \sum_{j\in \shifted}j^k \equiv 0 \bmod 3n^{k+1}+1.$$
Notice that $|\delta_{e_x}\cdot \sot^{\etag_x}j^k|<\sot^n j^k<n\cdot n^k=n^{k+1}$. This is also true for $|\delta_{\etag_y}\cdot \sot^{e_y}j^k|$ and $|\sum_{j\in \shifted}j^k|$.
As a result, the left part of the equivalences is at least $-3\cdot n^{k+1}$ and at most $3\cdot n^{k+1}$. Therefore, the congruences are strict equalities.

The equalities can be rewritten as
$$\sum_{j\in \shifted}j^k = -\delta_{\etag_y}\cdot(-\sot^{\etag_x}j^k + \sot^{e_y}j^k) = -\delta_{\etag_y}\cdot (\sum_{\etag_x+1}^{e_y}j^k).$$

Notice that the left hand side is always non-negative. Hence, the sign of the right hand side is non-negative as well.
From this follows that $\delta_{\etag_y}=-1$ and the equation can be transformed to
\begin{equation}\label{eq:basesums}
\sum_{\etag_x+1}^{e_y}j^k = \sum_{j\in \shifted}j^k.
\end{equation}

Since the current assumption is that $d_x<d_y$ and $ \etag_x<e_y$, there are 6 possible orderings of the 4 indices. 
A full proof for the cases $\etag_x,e_y<d_x$, $\etag_x<d_x<e_y<d_y$ will follow, and a guidance for the rest of the cases can be found afterwards.\par

Assume $\etag_x,e_y<d_x$. In this case, two sets are defined as
$$\sho \coloneqq [\etag_x+1,e_y], \, \sht \coloneqq \shifted.$$
Notice that for any two indices $i\in \sho,\ j\in \sht$ the following holds
\begin{equation}\label{eq:vandertwo}
i<j.
\end{equation}

Hence, \eqref{eq:basesums} can be altered to the following form. For any $k\in\{0,1,2\}$
$$\sum_{j\in\sho}j^k = \sum_{j\in \sht}j^k. $$

For $k=0$ this equality is $|\sho|=|\sht|$ which means the cardinality of the sets is equal.
For $k=1$ this equality is $\sum_{j\in\sho}j = \sum_{j\in \sht}j$  which means the sum of elements of $\sho,\sht$ is equal as well. However, through equality \eqref{eq:vandertwo} if the cardinality of the sets is the same then the sum of elements in $\sht$ should be strictly bigger than the sum of elements in $\sho$. This concludes this case.

Assume $\etag_x<d_x<e_y<d_y$. In this case, three sets are defined as{\small
$$\sho \coloneqq [\etag_x+1,d_x],\, \sht \coloneqq [d_x+1,e_y]\setminus \shifted,\, \shr \coloneqq[e_y+1,d_y]\cap \shifted$$}
Observe that for any three indices $i\in \sho,\ j\in \sht,\ \ell\in \shr$ the following holds
\begin{equation}\label{eq:vander}
i<j<\ell.
\end{equation}

Now, \eqref{eq:basesums} can be rewritten as follows.
For any $k\in\{0,1,2\}$
$$\sum_{j\in \sho}j^k +\sum_{j\in \sht}j^k-\sum_{j\in \shr}j^k=0.$$
This can be written in matrix form. 
There exist integers $v_1,v_2,v_3$ such that at least one of them is non-zero and the following equality holds
$$\bfA \cdot \bfv := 
\begin{pmatrix}
\sum_{j\in\sho} 1 & \sum_{j\in \sht}1 &\sum_{j\in \shr}1 \\
\sum_{j\in\sho} j & \sum_{j\in \sht}j& \sum_{j\in \shr}j\\
\sum_{j\in\sho} j^2 & \sum_{j\in \sht}j^2 &\sum_{j\in \shr}j^2
\end{pmatrix}
\cdot 
\begin{pmatrix}
v_1 \\
v_2 \\
v_3
\end{pmatrix} = 
\begin{pmatrix}
0 \\
0 \\
0
\end{pmatrix}
.
$$
In this case, $v_1=v_2=1,v_3=-1$ is such a solution. This equality means $\bfA$ has a non-trivial solution to the homogeneous system of equalities, which also means $\det(\bfA)=0$.

The determinant of the matrix $\bfA$ can be computed by
{\small$$
\begin{vmatrix}
\sum_{j\in\sho} 1 & \sum_{j\in \sht}1 &\sum_{j\in \shr}1 \\
\sum_{j\in\sho} j & \sum_{j\in \sht}j& \sum_{j\in \shr}j\\
\sum_{j\in\sho} j^2 & \sum_{j\in \sht}j^2 &\sum_{j\in \shr}j^2
\end{vmatrix}=
\sum_{i\in\sho} \sum_{j\in \sht} \sum_{k\in \shr} 
\begin{vmatrix}
1 & 1 & 1 \\
i & j & k \\
i^2 & j^2 & k^2
\end{vmatrix}.
$$}

Notice that each element in the sum is a determinant of a Vandermonde matrix. Hence,
$$
\begin{vmatrix}
1 & 1 & 1 \\
i & j & k \\
i^2 & j^2 & k^2
\end{vmatrix} = (j-i)\cdot(k-j)\cdot(k-i).
$$

To summarize, \eqref{eq:vander} the following contradiction is obtained.
\begin{flalign*}
0 &= \sum_{i\in\sho} \sum_{j\in \sht} \sum_{k\in \shr} 
\begin{vmatrix}
1 & 1 & 1 \\
i & j & k \\
i^2 & j^2 & k^2
\end{vmatrix} \\
&=\sum_{i\in\sho} \sum_{j\in \sht} \sum_{k\in \shr} 
 (j-i)\cdot(k-j)\cdot(k-i) >0.
\end{flalign*}
This concludes this case.

For any of the other orderings, the same proof can be concluded using the following definitions:
\begin{enumerate}
\item For $\etag_x,e_y<d_x$, define $\sho \coloneqq [\etag_x+1,e_y], \sht \coloneqq \shifted$;
\item For $\etag_x<d_x<e_y<d_y$, define $\sho \coloneqq [\etag_x+1,d_x], \sht\coloneqq[d_x+1,e_y]\setminus \shifted, \shr \coloneqq [e_y+1,d_y] \cap \shifted$;
\item For $\etag_x<d_x<d_y<e_y$, define $\sho \coloneqq [\etag_x+1,d_x], \sht \coloneqq \shifted, \shr \coloneqq [d_y+1,e_y]$;
\item For $d_x<\etag_x<e_y<d_y$, define $\sho \coloneqq \shifted\cap [d_x,\etag_x], \sht \coloneqq [\etag_x+1,e_y]\setminus \shifted, \shr \coloneqq [e_y+1,d_y]\cap\shifted$;
\item For $d_x<\etag_x<d_y<e_y$, define $\sho \coloneqq \shifted\cap [d_x,\etag_x], \sht \coloneqq [\etag_x+1,e_y]\setminus \shifted, \shr \coloneqq [d_y+1,e_y]$;
\item For $d_y<\etag_x,e_y$, define $\sho \coloneqq \shifted, \sht \coloneqq [\etag_x+1,e_y]$.
\end{enumerate}
This concludes the proof.
\end{IEEEproof}

In this proof, it is shown that $\cC_{a,b,c,d}$ guarantees to correct a combination of single-deletion and single-substitution errors. However, a single deletion or a single substitution can be corrected as well due to the $\vto$ constraint of the code $\cC(\vto;a,3n+1)$~\cite{Levenshtein-binarycodesCorrectingDeletions}. Lastly, via the pigeonhole principle the following conclusion about the redundancy is obtained.
\vspace{-2ex}

\begin{cor}
There exist $a\in [3n+1],b\in [3n^2+1],c\in [3n^3+1],d\in [5]$ such that the code $\cC_{a,b,c,d}$ is a binary single-deletion single-substitution correcting code with at most $6\cdot \log (n) +8$ redundancy bits.
\end{cor}


\section{Construction for Non-Binary Alphabets}\label{sec:qary}
In this section, a construction for non-binary codes correcting a single deletion and a single substitution is presented. 

For a word $\cww \in \Sigq^n$ we associate its \emph{binary signature} $\cww^{01} \in \Sigt^{n}$, where $\cww^{01}_1 = 1$ and $\cww^{01}_i = 1$ if and only if $\cww_i>\cww_{i-1}$.
The motivation for using the signature vector is to convert the error correction into a binary problem. 
The following lemma shows the conversion explicitly. We define an \emph{adjacent transposition} to be the error event in which two adjacent bits switch their positions.\vspace{-2ex}
\begin{lemma}
For any word $\cwx\in \Sigq^n$, and $\cwy\in \Sigq^{n-1}$ the error word achieved by a single-deletion and a single-substitution,
$\cwy^{01}	$ can be achieved from by $\cwx^{01}$ by one of the following errors:
\begin{enumerate}
\item single-deletion;
\item single-deletion and a single-substitution;
\item single-deletion and a single-adjacent-transposition.
\end{enumerate}
\end{lemma}

For the rest of this section, let $\cC_2$ be a code correcting either a single deletion and a single substitution or a single deletion and a single adjacent transposition. We are now ready to present the code construction for the non-binary case. \vspace{-1ex}
\begin{construction} 
For $a \in [2q], b\in [2qn]$, $c\in [2qn^2]$, let $\cC_{a,b,c}$ be the  code
{\small
$$\cC_{a,b,c}=\hspace{-0.5ex}\left\{\cwx\in\Sigq^n| \cwx^{01}\in \cC_2,\hspace{-0.5ex}
\begin{array}{ll}
\sot^n x_j \equiv \hspace{-0.5ex} a (\bmod\ 2\cdot q+1),\\
\sot^n j\cdot x_j \equiv \hspace{-0.5ex} b (\bmod\ 2\cdot n\cdot q+1),\\
\sot^n j^2\cdot x_j \equiv \hspace{-0.5ex} c (\bmod\ 2\cdot n ^2 \cdot q+1)
\end{array} 
\hspace{-1ex}
 \right\}$$
}
\end{construction}

The next theorem states the correctness of this code construction. \vspace{-2ex}
\begin{theorem}
For all $a \in [2q], b\in [2qn]$, $c\in [2qn^2]$ the code $\cC_{a,b,c}\subseteq \Sigq^n$ is a single-deletion single-substitution correcting code.
\end{theorem}\vspace{-1ex}

Notice that this construction requires an extension of the binary code presented in Section \ref{sec:cons}. 
Such an extension is possible using the result presented in \cite{Gabrys_Codes-in-the-Damerau-Distance-for-Deletion-and-Adjacent-Transposition-Correction_18} for a family of binary codes correcting a single deletion and a single adjacent-transposition. Hence, it is possible to conclude with the following corollary.\vspace{-1ex}
\begin{corollary}
There exist $a \in [2q], b\in [2qn]$, $c\in [2qn^2]$ such that the code $\cC_{a,b,c}$ is a single-deletion single-substitution code with at most $10\cdot \log(n) +3\cdot \log(q)+11$ redundancy symbols.
\end{corollary}

\bibliographystyle{IEEEtran}
\bibliography{refs}

\end{document}